\documentclass[aps,prl,twocolumn,groupedaddress,nofootinbib,superscriptaddress]{revtex4}

\usepackage{amsmath,amssymb,epsfig,ifthen,capt-of,calc}

\newcommand{\bignone}{}
\newcommand{\mathd}{\mathrm{d}}
\newcommand{\mathpi}{\pi}
\newcommand{\tmem}[1]{{\em #1\/}}
\newcommand{\tmop}[1]{\ensuremath{\operatorname{#1}}}
\newcommand{\tmstrong}[1]{\textbf{#1}}
\newcommand{\tmtextit}[1]{{\itshape{#1}}}

\begin{document}


\title{A Negative S parameter from Holographic Technicolor}



\author{Johannes Hirn}
\affiliation{IFIC -  Universitat de Val\`encia, Edifici d'Instituts de Paterna,  Apt. Correus 22085, 46071 Val\`encia, Spain}

\author{Ver\'onica Sanz}
\affiliation{Departamento de F\'\i sica Te\'orica y del Cosmos,  Universidad de Granada, Campus de Fuentenueva, 18071 Granada,  Spain}


\begin{abstract}
  We present a new class of 5D models, Holographic Technicolor, which fulfills
  the basic requirements for a candidate of comprehensible 4D strong dynamics
  at the electroweak scale. It is the first Technicolor-like model able to
  provide a vanishing or even negative tree-level $S$ parameter, avoiding any
  \tmtextit{no-go theorem} on its sign. The model is described in the
  large-$N$ regime. $S$ is therefore computable: possible corrections coming
  from boundary terms follow the $1 / N$ suppression, and generation of
  fermion masses and the $S$ parameter issue do split up. We investigate the
  model's 4D dual, probably walking Technicolor-like with a large anomalous dimension.
  \end{abstract}

\maketitle


{\bf Introduction: }  
The idea that electroweak symmetry breaking (EWSB) could be due to the onset of the strong-coupling regime in
an asymptotically-free gauge theory was first put forward to solve the
hierarchy problem in  \cite{Weinberg:1979bn}. Technicolor was based on the example of
massless QCD with two flavors, where the global $\tmop{SU} (2)
\times \tmop{SU} (2)$ symmetry is spontaneously broken to the diagonal
subgroup. A similar
theory with a mass scale of order $3000$ larger would feed its three GBs to
the SM $\tmop{SU} (2)_L \times \text{U} (1)_Y$ gauge fields, yielding masses
for the $W^{\pm}$ and $Z$, without an associated Higgs boson. It was however
shown that a simple rescaled version of QCD fails, since it leads to the
famous $S$ parameter being too large and positive as compared to the value
extracted from experiments \cite{Holdom:1990tc}, unless the number of techni-colors is
small. This last possibility is however undesirable, as it signifies the loss
of our last non-perturbative handle, namely the large-$N$ expansion.

The recent developments in Holographic QCD \cite{Erlich:2005qh,daRold:2005zs,Hirn:2005nr}  give us a computable way
of departing from rescaled QCD. The models of Holographic QCD aim to describe
the dynamics of the QCD bound states in terms of a 5D gauge theory: the input
parameters in such a description can be identified with the number of colors,
the confinement scale and the {\tmem{condensates}}. The present class of
models for dynamical EWSB works in a similar spirit. For the first time, the
tree-level $S$ parameter is negative. This has further consequences in the
gauge boson spectrum.

{\bf Holographic Technicolor: }
Our starting point is a model in five-dimensions (5D) describing electroweak
  symmetry breaking via boundary conditions (BCs). The extra dimension we
  consider here is an interval. The two ends of the space are located at
  $l_0$ (the {\tmem{UV brane}}) and $l_1$ (the {\tmem{IR brane}}), with the
  names UV/IR implying $w \left( l_0 \right) \geqslant w \left( l_1 \right)$.
  We focus on  metrics that can be recast as
 $   \mathd s^2 = w \left( z \right)^2  \left( \eta_{\mu \nu} \mathd
    x^{\mu} \mathd x^{\nu} - \mathd z^2 \right) $.  We only consider the dynamics of the bulk 5D symmetry $\tmop{SU} (2)_L
  \times \tmop{SU} (2)_R \times U (1)_{B - L}$ gauge symmetry
. As in Higgsless models
  {\cite{Csaki:2003dt}}, the BCs are
  chosen to break the LR symmetry to the diagonal $\tmop{SU} (2)_D$ on the IR
  brane, while the breaking on the UV brane reduces $\tmop{SU} (2)_R \times U
  (1)_{B - L}$ to the hypercharge subgroup. The remaining 4D gauge symmetry is
  thus $U (1)_Q$.

  An important ingredient of Holographic Technicolor comes from the
  lessons learned in Holographic QCD: breaking on the brane is too soft to account for all phenomena
  found in  QCD, in particular power corrections at high energies due to
  condensates. Besides this breaking by BCs, we therefore introduce breaking in the bulk. In the following, the bulk source of
  breaking will be a crossed kinetic term between L and R gauge fields, just
  as in {\cite{Hirn:2005vk}}. (The $z$-dependence of this term could be
  obtained from the profile of a scalar.) At the quadratic level, this
  well-defined procedure may {\it effectively} be
  summarized  as yielding different metrics, $w_A(z) \neq
  w_V(z)$ {\cite{Hirn:2005vk}}. This bulk breaking will play an important role in
  our description of strong dynamics at the TeV.
\\
\,
\\
{\bf The spectrum: } In terms of physical states, no massless mode
survives except
  for the photon. The remainder will pick up masses via the compactification. For the
  class of metrics that decrease away from the UV as AdS or faster ({\tmem{gap
  metrics}}), the massive modes can be separated into two groups:
  {\tmem{ultra-light excitations}} {\cite{Csaki:2003dt}} and
  {\tmem{KK-modes}}. If we interpret the ultra-light modes as the $W$
  and $Z$, the gap suppresses the KK contributions to the electroweak
  observables
  {\cite{Csaki:2003dt}}: this can be seen clearly using Sum
  Rules (SRs).

For any gap metric, the KK modes are repelled from the UV brane, and the massive modes approximately
  split into separate towers of axial and vector fields (and $B$ fields).
  Thus, $W'$, the first KK mode above the $W$ would {\tmem{a priori}} be a
  vector (the techni-rho), while the next one, $W''$, would be an axial
  resonance (techni-$a_1$), etc... One can extract SRs involving KK-mode
  masses (excluding ultra-light modes)
  \begin{eqnarray}
    \sum_{n = 1}^{\infty} \frac{1}{M_{X_n}^2}    & \simeq & \int_{l_0}^{l_1} \mathd zw_X \left( z \right) \alpha_X \left( z
    \right)  \int_{l_0}^z \frac{\mathd z'}{w_X \left( z' \right)}  ,  \label{SRWp}
  \end{eqnarray}
 where $X = V, A, B$ and $\alpha_{V, B} \left( z \right) = 1$
  and $\alpha_A \left( z \right) = \int_z^{l_1} \frac{\mathd z'}{w_A \left( z'
  \right)} / \int_{l_0}^{l_1} \frac{\mathd z''}{w_A \left( z'' \right)}$. The
  SR in Eq.(\ref{SRWp}) is exact at order
  $\mathcal{O}\left(G^0\right)$, where $G$ is the gap between the ultra-light mode and the heavy modes: in AdS,
  $w (z) = l_0 / z$ and the gap is $G = \log (l_1 / l_0)$. As in Holographic QCD, the
  function $\alpha_A (z)$ {\cite{Hirn:2005nr}} is the wavefunction of the
  ``would-be'' Goldstone boson matrix, $D_{\mu} U (x)$: it is monotonously decreasing with
  BCs $\alpha_A \left( l_0 \right) = 1$ and $\alpha_A \left( l_1 \right) =
  0$. \

\begin{figure}
  \includegraphics[width=8cm]{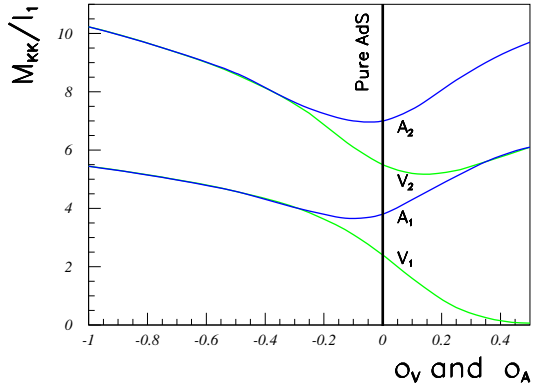}
 \caption{\label{spectrum}Masses at $\mathcal{O}\left(G^0\right)$
divided by $l_1$ for the lightest vector and axial KK
 modes of the $W$, as a function of the condensate in their
 respective channel  $o_{V,A}$, for $d=2$.}
  \end{figure}

  On the other hand, another exact SR can be obtained, involving both heavy
  {\tmem{and}} ultra-light modes. For gap metrics, it can be expanded to
  obtain the mass of the ultra-light mode: at
  order $\mathcal{O}\left(1/G\right)$, we get
  \begin{eqnarray}
    M_W^2 & \simeq & 1 / \left(   \int_{l_0}^{l_1} \mathd z (w_V (z) + w_A (z))  \int_{l_0}^{l_1} \frac{\mathd z'}{w_A \left( z'
    \right)}\right),  \label{SRW}
  \end{eqnarray}
  which can be shown to agree with the expression involving the 4D gauge
  coupling $g$ and techni-pion decay constant $f$
  \begin{eqnarray}
    M_W^2 & = & \frac{g^2 f^2}{4} +\mathcal{O} \left( 1 / G^2 \right) =
    \frac{1}{Gl_1^2} +\mathcal{O} \left( 1 / G^2 \right) , \label{MWgf}
  \end{eqnarray}
as expected from Technicolor. Eq.(\ref{MWgf}) shows that, at leading order $\mathcal{O}\left(1/G\right)$,
  $M^2_{W, Z}$ do not feel  any breaking of conformality in the bulk: their mass is
  dominated by the UV physics. On the other hand, Eq.(\ref{SRWp}) showed that
  the KK masses do feel the effect of this bulk breaking at leading
  order $\mathcal{O}\left(G^0\right)$. Also, since their
  wave-functions are repelled, the
  KK-modes have masses that are quite insensitive to the UV brane
  position. Their
  appearance is due to the fragmentation of the continuum of states due to the IR
  breaking of conformal invariance. Therefore, their mass is dictated by the
  position of the IR brane, $m \propto 1 / l_1$ and does in addition depend on the
  condensates, as shown in FIG. \ref{spectrum} for the metrics of Eq.(\ref{metric2}): the ratio $m_{A_1} / m_{V_1}$
  tends to be lowered as negative condensates are switched on.

  Many other results can be shown in terms of SRs {\cite{usprep}}, and we
  just outline them briefly. For example, as is standard in 5D models,
  non-oblique corrections are produced at low-energy: four-fermion
  interactions are generated by the exchange of KK states. It can be shown that the expression of
  the resulting Fermi constant in terms of the techni-pion decay constant is
  obtained from the SM by replacing $v \rightarrow f$. Since the model is
  based on an $\tmop{SU} (2) \times \tmop{SU} (2)$ (gauge) symmetry in the
  bulk, broken to the diagonal subgroup by the IR BCs, it possesses custodial
  symmetry. This implies that the low-energy rho parameter
  $\rho_{\ast} \left( 0 \right)$ is strictly equal to one at tree level, as
  was found in the deconstructed case
  {\cite{Hirn:2004ze}}
  and indicated by {\cite{Georgi:1978wk}}. Also, the KK modes, being repelled
  from the UV brane, are insensitive to the UV BCs. The KK spectrum is
  therefore isospin symmetric up to $1 / G$ corrections: $W_n$ degenerate with
  $Z_n$. In addition, since the KK contribution is small due to the large
  masses of the KK modes, one concludes that the $T$ parameter is suppressed in these models. Finally, one can
  also show from two SRs that the $E^4$ and $E^2$ contributions to the $W_L
  W_L $ scattering vanish {\cite{Csaki:2003dt}}.
\\
\,
\\
  {\bf The $S$ Parameter: }
  The tree-level contribution to the $S$ parameter, being a
  low-energy effect due to strong dynamics responsible for spontaneous
  symmetry-breaking, can be expressed {\cite{Holdom:1990tc}} in
  terms of the $L_{10}$ coupling of chiral lagrangians
 $    S_{\tmop{tree}}  =  - 16 \mathpi L_{10} $.
  The value extracted from LEP physics is
  {\cite{Eidelman:2004wy}} $S = - 0.13 (0.07) \pm 0.10$ with
  reference Higgs mass $m_H = 117 (150) \tmop{GeV}$, where the value in parentheses
  is the most recent analysis of data at the $Z$ pole (2005). A sizeable negative $L_{10}$ would easily upset the experimental constraint  (note that in $N_c=3$ QCD,  $- 16 \mathpi L_{10} \sim 0.3 $).
On the other hand,  large-$N$ models of strong dynamics
  predict the value of $L_{10}$ in terms of contribution of spin-1 resonances
 $   L_{10}   = - 1/4 \sum_{n = 1}^{\infty} f_{V_n}^2 - f_{A_n}^2$
,  via their decay constants $f_{X_n}$ according to {\cite{Ecker:1989yg}},
whereas other
  contributions are down by $1 / N$.  Higgsless models thus face a serious challenge, a \tmtextit{no-go theorem}
  {\cite{Barbieri:2003pr}}: $L_{10}$ is bounded to be
  negative. This is readily understood by using a SR: one can translate the
  sum over resonance contributions into a purely geometric factor
  \begin{eqnarray}
    L_{10} & \bignone = & - \frac{N}{48 \pi^2}  \int_{l_0}^{l_1} \frac{\mathd z}{l_0} w
    (z) \left( 1 - \alpha (z)^2 \right),  \label{L10BC}
  \end{eqnarray}
  where we have defined $N / 12 \pi^2 \equiv  l_0 / g_5^2$. The
  bound $\alpha \left( z \right) \leqslant 1$ implies that $L_{10}$ is
  negative and proportional to the loop expansion parameter, $N$. The most
  natural value for $L_{10}$ will thus drive a large  positve $S$ parameter,
  excluding the simplest realization of the model. For example, pure AdS yields
  $S_{\text{tree}} = N / 4 \pi$.

  One possibility would be to consider these models in the low-$N$ regime.
  This situation is most unwelcome, as has been stressed by many authors
  {\cite{Luty:2004ye}}
The main reason is that the value
  of $N$ plays an important role: it sets the range of computability of the
  model. Low $N$ implies strong coupling of the gauge KK modes. A way
  of putting it is via the \tmtextit{position-dependent cutoff
  }{\cite{Randall:2001gc,Randall:2002tg}}: a cutoff  $\Lambda$ at the position where $w \left(z\right)$ is normalized to unity will be redshifted for
  processes located near a position $z$ as $\Lambda (z) = \Lambda
  \sqrt{g_{00}} = \Lambda w (z)$. For example, in pure AdS, the 5D loop
  expansion breaks down when $\Lambda (z) z \sim 24 \mathpi^3 l_0 / g_5^2 = 2
  \mathpi N$. The other parameter playing an important role is the gap $G.$
  Reproducing the Fermi constant and the $W$ mass
  implies $N G \sim 500$. Pushing to low values of $N$ is thus asking for a bigger
  separation between the $W$ and its KK modes, which would conflict with
  the premise of perturbativity: strong coupling would set in before the resonances tame the high-energy behavior of amplitudes.

  Returning to  the large-$N$ regime, one is then cornered to hope for
  miraculous cancellations. Efficient possibilities would be:
  introducing IR localized
  kinetic terms proportional to $\tmop{SU} (2)_D$ or hoping for cancellations
  against fermion contributions
  {\cite{Csaki:2003dt}}. Both possibilities face
  again new challenges, difficult to resolve. Trying to add large
  localized kinetic terms with the ``wrong'' sign, which are of order $1 / N$ directs again towards
  the low-$N$ problem. Besides it leads to a tachyon instability
  {\cite{Barbieri:2003pr}}. The way out with bulk fermions poses a problem of
  naturalness and dangerously ties the $S$ parameter problem with the fermion
  mass hierarchy, and therefore with non-oblique corrections
  {\cite{Csaki:2003dt,Georgi:2005dm}}.

  Here we propose a different point of view, which arises naturally in Holographic QCD
  and should therefore appear in a Technicolor-like model. Local order
  parameters of the symmetry-breaking imply a \tmtextit{different} behavior
  for the $V$ and A combinations of bulk fields
  {\cite{Erlich:2005qh,Hirn:2005nr}}. In the simplest
  realization of this IR behavior {\cite{Hirn:2005vk}}, $L_{10}$ is modified
  from Eq.(\ref{L10BC}) to read
  \begin{eqnarray}
    L_{10} & = & - \frac{N}{48 \pi^2}  \int_{l_0}^{l_1} \frac{\mathd z}{l_0} \left( w_V
    (z) - w_A \left( z \right) \alpha_A (z)^2 \right),  \label{L10bulk}
  \end{eqnarray}
  where $w_{V, A}$ are the metrics felt by the axial and vector combinations
  of fields.

  $L_{10}$ is still \tmtextit{proportional to N}, but the integrand in
  Eq.(\ref{L10bulk}) can reverse sign for $z$ such that $w_A \left( z \right)
  \alpha \left( z \right)^2 > w_V \left( z \right)$, and  $L_{10}$ may come
  out positive. The first consequence is quite clear: a large-$N$ scenario is
  then preferred, extending the pertubativity regime. In particular, the bulk
  value of $S$ will {\tmem{not}} receive sizeable corrections from the
  localized kinetic terms, since these are still suppressed by $1 / N$. The
  $S$ parameter is therefore computable. Another important property of the
  bulk $S$ parameter is its independence on the exact IR dynamics. Contrary
  to the spectrum, contributions to $S$ come mainly from the bulk far from
  the branes {\cite{usprep}}.

  We now assume that the metrics behave as AdS near the UV brane and deviate from conformality in the bulk according to
  \begin{eqnarray}
    w_X \left( z \right)  & = &  \frac{l_0}{z} \exp \left( \nu_X  \left(
    \frac{z - l_0}{l_1 - l_0} \right)^{2 d} \right) .  \label{metric2}
  \end{eqnarray}
  As explained at the beginning of the paper, this parametrically simple
  form encodes effects of couplings with other background fields, whose
  dynamics we neglect here. At order $\mathcal{O}\left(G^0\right)$,  one can obtain an analytic expression for $S$ in
  the case $\nu_A = 0$ and $\nu_V < 0$
  \begin{eqnarray}
    S_{\text{tree}} & = & \frac{N}{4 \pi}  \left( 1 - \frac{2}{3 d} (\Gamma (-
    \nu_V) + \log (- \nu_V) + \gamma_E) \right) . \label{SVanalytic}
  \end{eqnarray}
  In $\nu_V \equiv \frac{12 \pi^{3 / 2} \Gamma (d + 1 / 2)}{d^2 \Gamma
  (d)^3} o_V$, NDA sets $o_V \sim \mathcal{O}(1)$
  {\cite{Hirn:2005vk}}.

  \begin{figure}
  \includegraphics[width=8cm]{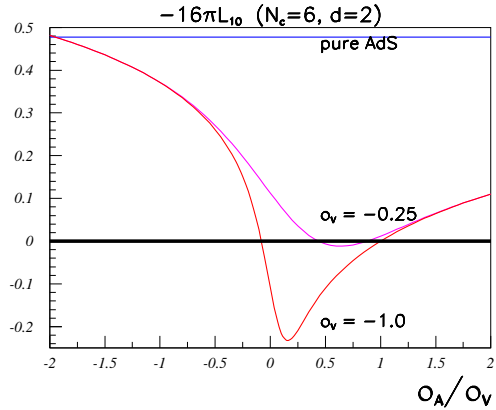}
 \caption{\label{srat}Value of $S_{\tmop{tree}}/N$ ---for
 $d=2$ and for different values of $o_V$--- as a function of the ratio of condensates in the two channels
 $o_A/o_V$, and for the pure AdS case. }
  \end{figure}

 In FIG.(\ref{srat}),
  we show the value of $S_{\tmop{tree}}/N$ for different values of the ratio $o_A / o_V$
  fixing $d = 2$. A negative vector
  condensate can lead to vanishing or negative $S_{\tmop{tree}}$, even more so if it is accompanied by an axial condensate of the same sign: a direction not explored by the authors of \cite{daRold:2005zs,Hong:2006si}. Also,
  assuming $o_X \sim \mathcal{O} \left( 1 \right)$, the effect disappears if the
  dimension of the condensate is increased. Our results thus extend those of
  {\cite{Sundrum:1991rf}}, which indicated
  that increasing $o_A - o_V$, preferably with a low $d$, could decrease
  the $S$ parameter, in connection with a lowering of the ratio $m_{A_1}/m_{V_1}$.

  A refinement in the computation of the $S$ parameter comes from taking into
  account the pion loop effects {\cite{Holdom:1990tc}} and subtracting the SM
  value with a reference Higgs mass
  \begin{eqnarray}
    S & = & - 16 \mathpi L_{10} \left( \mu \right) + \frac{1}{12 \mathpi}
    \left( \log \left( \frac{\mu^2}{m_H^2} \right) - \frac{1}{6} \right) .
    \label{Sloop}
  \end{eqnarray}
  From the understanding of the QCD case {\cite{Ecker:1989yg}}, one expects
  the model to predict the value of $L_{10} \left( \mu \right)$ at the
  matching scale of the model with a chiral lagrangian, i.e. $\mu \sim
  \text{few}/l_1 \sim \text{few}\, \tmop{TeV}$, the mass scale of the resonances. The second term in
  Eq.(\ref{Sloop}) is then positive and of order $0.1$, requiring a vanishing or
  slightly negative $S_{\tmop{tree}}$, as provided by the present model.
\\
\,
\\
  {\bf Four-dimensional dual: }
  Holographic models are inspired from the AdS/CFT correspondence
  {\cite{Gubser:1998bc}}. The precise form of this conjecture
  relates two highly symmetric theories and is, unfortunately, far from being
  of direct phenomenological relevance. After a pioneering work by Pomarol
  {\cite{Pomarol:2000hp}}, authors in {\cite{Arkani-Hamed:2000ds}} explored
  the audacious conjecture that more realistic models like Randall-Sundrum
  {\cite{Randall:1999ee}} would somehow inherit properties of the duality.
  Since then, more evidence has been gathered towards a
  5D/4D duality, the latest being bottom-up models of Holographic QCD \cite{Son:2003et,Erlich:2005qh,daRold:2005zs,Hirn:2005nr,Hirn:2005vk}. The success of these models in capturing the
  behavior of a strongly-coupled theory like QCD provides an incentive
  for applications to Technicolor. In this case, one starts off on a firmer footing:
  in the presence of condensates, the number of (techni)-colors can be made large since
  it no longer in conflict with the $S$ parameter.

  Let us show the effect in the 4D two-point correlator of the current $X = V,
  A, B$ of a metric of the form given by Eq.(\ref{metric2}). For large euclidean $Q^2$, the two-point
  function for this field $X$ reads {\cite{Hirn:2005nr}}
  \begin{eqnarray}
    \Pi_X \left( - Q^2 \right)    & \simeq & - \frac{N}{12 \mathpi^2}  \left( \log \left( \frac{Q^2}{\mu^2}
    \right) + \lambda \left( \mu \right) \right) + \frac{\left\langle
    \mathcal{O}_{2 d_X} \right\rangle}{Q^{2 d_X}}   \label{PiX}
  \end{eqnarray}
  where the parameter $o_X \equiv \left\langle \mathcal{O}_{2 d_X}
  \right\rangle / (Nl_1^{- 2 d_X}) \sim \mathcal{O}(1)$. To have a chance of
  obtaining a positive value for $L_{10}$, we need $\left\langle O_{2 d}
  \right\rangle_V < \left\langle O_{2 d} \right\rangle_A$. This is in
  agreement with Witten's positivity condition for $\Pi_A - \Pi_V$
  {\cite{Witten:1983ut}}, ensuring the stability of the selected vacuum
  {\cite{Dashen:1971et}}.
  Holography tells us that this bulk field $X$ is dual to some operator
  $\mathcal{O}$ on the 4D side with the same quantum numbers: the correlators
  generated by $X$ and by $\mathcal{O}$ are the same. In this particular case
  we see that deviations from conformality with a given power of $z^{2 d}$ in
  Eq.(\ref{metric2}) mimick the effects of a condensate of dimension $2 d$ in
  the 4D dual.

  Generally speaking, non-perturbative effects in QCD-like Technicolor models
  make them unreliable. The same goes for the case of a flat extra-dimension,
  the cutoff of the theory is quite low, $\Lambda \sim 2 \mathpi N /
  l_1${\tmstrong{ }}and quantities like the $S$ parameter are no longer
  computable. On the other hand, extra-dimensional models in AdS behave
  in a similar fashion to walking Technicolor. The warping suppresses convolutions of
  wave-functions, as walking kills unwanted operators. But in pure AdS, one
  cannot choose which operators will be suppressed: their scaling is dictated
  by the warping, whereas gap metrics with violations of conformality like
  Eq.(\ref{metric2}) do change the scaling.

  The dual of Holographic Technicolor must be a strongly-coupled theory, with the running in the
  UV dictated by the one of a gap metric and with non-perturbative dynamics
  affecting the vector and axial channel in a similar way.   If the 4D dual is going to yield small or negative $S$ parameter, the net
  effect of condensates in the vector and axial current must go in the
  direction of $w_A \alpha^2 > w_V$.
  For example, imagine
  that strong dynamics generate a techni-condensate $\langle Q \bar{Q} \rangle$
  responsible of breaking the Technicolor gauge group $\tmop{SU} (N)$: this
  condensate is represented in the 5D dual as the rescaled vev of $\langle
  \Phi \rangle$. Assume now that the anomalous dimensions is large, for example, due to the running
  mass in the 5D picture. Then, there will be a difference between the
  canonical dimension of $\langle \bar{Q} Q \rangle$ and the running dimension
  of the operator. A way of modelling this anomalous dimension would be that the vector and axial fields
 couple to a scalar representing the techni-quark condensate,
  $\Phi$, via a{\tmem{ running mass}}, such that $m_{\Phi} (l_0)^2 = - 3 /
  l_0^2 $ and $m_{\Phi} (l_1)^2 = d (d - 4) / l_0^2$ with $d<3$
  ($d=2$ for extreme walking).
\\
\,
\\
  {\bf Conclusions: }
In this paper we have shown quantitatively how technicolor models
which depart from
rescaled QCD can exhibit a negative tree-level $S$ parameter. This was done
using a holographic model (i.e. using a 5D gauge theory) for the resonances
created by a strongly-interacting theory such as technicolor. It is based on
the recent successes of similar 5D models for the resonances of QCD. These
successes themselves validated the idea of the duality between 4D
strongly-coupled theories and 5D weakly-coupled ones at the quantitative
level.

  We have presented the first Technicolor-like model able to provide a small
  $S$ parameter, and to remain computable since it is defined in the large-$N$
  limit. The 5D picture shows generic features of this class of models: 1) the
  metric has to fall off fast near the UV to generate a gap, 2) deviations
  from conformality must be introduced in the bulk, describing condensates, 3)
  a condensate of natural size can produce the desired effect if it has
  dimension close to 4 (as would happen for $\alpha_{\tmop{TC}} \langle \bar{Q} Q \rangle^2$ in walking Technicolor), 4) $W'$ and $Z'$ (vector resonances) then tend to
  become degenerate with the $W''$ and $Z''$ (axial) resonances.

 In the present paper, the fermions were located for simplicity on the UV brane.  As soon as we let them live in the bulk, much more interesting phenomena
  should arise: one big advantage of the present models is that the fermion
  profiles are not constrained by the requirement of cancelling  the
  $S$ parameter contributions. The issue of the $S$ parameter is therefore decoupled from that of fermion mass generation or from $Z \rightarrow b
  \overline{b}$, which can be addressed in a new view {\cite{usprep}}. In particular,
  topcolor assisted models would be implemented as in
  {\cite{Rius:2001dd}}.
  {\bf Acknowledgments: }
We acknowledge hospitality from Boston, Harvard and Yale Universities
during the completion of this work. We also thank Tom Appelquist, Tony
Gherghetta, Ami Katz, Ken Lane, John March-Russell, Toni Pich and Francesco
Sannino for stimulating discussions. JH is supported by the EC RTN
network HPRN-CT-2002-00311 and by the Generalitat Valenciana grant GV05/015.

\bibliography{bib-biblio}

\end{document}